\documentclass[prl,aps,twocolumn,showpacs]{revtex4}
\usepackage{graphicx}
\usepackage{amsmath}
\usepackage{bm}
\begin{document}

\title{Scattering Theory for Quantum Hall Anyons in a Saddle Point Potential} 

\author{A. Matthews}
\affiliation{T.C.M. Group, Cavendish Laboratory, J.~J.~Thomson Ave., Cambridge CB3~0HE, UK.}

\author{N. R. Cooper}
\affiliation{T.C.M. Group, Cavendish Laboratory, J.~J.~Thomson Ave., Cambridge CB3~0HE, UK.}

\begin{abstract} 

  We study the theory of scattering of two anyons in the presence of a
  quadratic saddle-point potential and a perpendicular magnetic
  field. The scattering problem decouples in the centre-of-mass and
  the relative coordinates.  The scattering theory for the relative
  coordinate encodes the effects of anyon statistics in the
  two-particle scattering. This is fully characterized by two
  energy-dependent scattering phase shifts.  We develop a method to
  solve this scattering problem numerically, using a generalized
  lowest Landau level approximation.

\end{abstract} 

\date{September 11, 2009}
\maketitle

%\section{Introduction}

One of the most remarkable features of the fractional quantum Hall
effect\cite{prangeandgirvin,dassarmapinczuk} is the existence of
quasiparticles with fractional charge\cite{laughlinstates} and
fractional exchange statistics\cite{halperinhierarchy,arovas}.  Direct
evidence for the existence of fractional charge has been observed
experimentally\cite{goldmancharge,FRACTION,glattlicharge,yacobycharge}. One
way this has been achieved is through the study of shot noise in a
quantum Hall system constricted by a point
contact\cite{FRACTION,glattlicharge}, by which the charge of the
current carriers may be derived from the fluctuations of the current
backscattered from the point contact.

Despite strong theoretical reasons to expect that these
fractionally charged quasiparticles also have fractional
statistics\cite{halperinhierarchy,arovas}, to date there has been no
unambiguous experimental demonstration of fractional statistics.  Several
authors have proposed methods in which evidence of fractional
statistics might appear in transport
experiments\cite{PhysRevLett.83.580,PhysRevLett.91.196803,TRIANGLE,law:045319,feldman:085333},
based on mesoscopic devices of various geometries.  Here we adopt a
simple approach, and ask a very natural theoretical question: can
anyonic statistics influence the transport through a {\it single}
point contact?  Clearly, since the effect of exchange statistics is a
two-particle property, any such effect requires 
more than one particle to be present in the point contact region.  We
therefore study the scattering properties of a pair of anyons incident
on the point contact.  Since the quantum Hall anyons are charged, and
experience a strong magnetic field, their free motion is along edge
states which follow classical equipotentials. The required scattering
theory is therefore very different from that of freely moving anyons,
described by a conventional kinetic energy\cite{korff-1999-40}.
As we show, the anyonic nature does affect the scattering of two
quantum Hall quasiparticles in the point-contact region.  This effect
can be characterized by two energy-dependent phase shifts, which we
calculate numerically for general anyonic statistics parameter.

%\section{Theoretical Model}

We consider two identical particles, with mass $M$ and charge $q$,
subjected to a magnetic field $B$ and in the presence of a quadratic
saddle point potential
\begin{equation}
H= \sum_{i=1,2}
\frac{1}{2M}\left[ \bm{p}_i -q \bm{A}(\bm{r}_i) \right]^2 + U\left(y_{i}^2  - x_{i}^2\right)
\label{eq:ham}
\end{equation}
where $\nabla\times \bm{A} = B\hat{\bm{z}}$ is the uniform magnetic
field.  
The particles are taken to be anyons\cite{wilczek}, with statistics parameter
$\Delta$. Thus, the two-particle wave function has the boundary
condition that
\begin{equation}
\Psi \left(\theta+2\pi\right)=e^{-i2\pi\Delta}\Psi  \left(\theta\right) 
\label{eq:anyon}
\end{equation}
where $\theta \to \theta + 2\pi$ represents one complete clockwise
rotation of the relative co-ordinate of the two particles.
We note that this defining feature applies either for
distinguishable or indistinguishable anyons\cite{footnote}.  If the anyons are
indistinguishable, one can impose the stronger restriction
\begin{equation}
\Psi \left(\theta+\pi\right)=  e^{-i\pi\Delta}\Psi  \left(\theta\right) 
\quad
\mbox{[indistinguishable]}
\label{eq:indistinguishable}
\end{equation}
Then, one recovers the exchange statistics for bosons and fermions for
$\Delta = 0, 1$ respectively. For greater generality, we study the
case of distinguishable anyons. The effects of the additional
restriction for indistinguishability (\ref{eq:indistinguishable})
will be made clear in the discussion.
 
We simplify the problem, defined by (\ref{eq:ham}) and the boundary condition (\ref{eq:anyon}), by introducing a ``centre-of-mass''
coordinate, $ \bm{r}_{c} $, and a ``relative'' coordinate, $ \bm{r}_{r} $, with
\begin{equation}
\bm{r}_{c}\equiv \frac{1}{\sqrt{2}}\left(\bm{r}_{1} + \bm{r}_{2}\right) \quad
\bm{r}_{r}  \equiv \frac{1}{\sqrt{2}}\left(\bm{r}_{1} - \bm{r}_{2}\right) 
\end{equation}
The Hamiltonian separates, becoming
\begin{equation}
H_\alpha \equiv  \sum_{\mu = c,r} \frac{1}{2M}\left[ \bm{p}_\mu -q \bm{A} \left(\bm{r}_\mu\right) \right]^2  + U\left(y_\mu^2  - x_\mu^2\right)
\end{equation}
 Thus, the two-particle problem can be expressed
as two independent one-particle problems, and the total energy $E =
E_c + E_r$ can be divided into separately conserved contributions from
the centre-of-mass and relative co-ordinates.

The centre-of-mass co-ordinate $\bm{r}_{c}$ is insensitive to anyonic
statistics. The scattering theory for this co-ordinate is identical to
the one-particle problem solved by Fertig and Halperin\cite{FANDH}.
This work allows one to deduce the transmission and reflection
coefficients of the centre-of-mass co-ordinate moving in the saddle
point potential, in terms of $E_c$.

The relative coordinate $ \bm{r}_{r} $ has the same
Hamiltonian. However, since the wave function has the additional
anyonic boundary condition (\ref{eq:anyon}), the results of
Ref.\cite{FANDH} cannot be applied. The solution of the scattering
problem for the relative co-ordinate is the central result of the
present paper.  From here on, for simplicity, we drop the subscripts
$r$ on $\bm{r}_r$ and $E_r$, with the understanding that the
calculation refers only to the relative co-ordinate.

%\section{Generalized Landau Level Spectrum}

Before considering the scattering problem, we study first the spectrum
for the relative co-ordinate in the absence of a potential, $U=0$. For
$\Delta =0$, this leads to the familiar Landau level states and
spectrum.  The anyon boundary condition changes the nature of these
states. To 
describe the generalized Landau level states, we  use polar coordinates,
$\bm{r} = r(\sin \theta, \cos \theta )$
and
the symmetric gauge
$\bm{A}=-\bm{e}_{\theta} rB/2$.
The eigenstates are
\begin{equation}
\psi_{n,m}(r,\theta) =e^{-i(m+\Delta)\theta}\;R_{n,m}\left(r\right)
\end{equation}
where $m$ is an integer, and $n=0,1,2,\ldots$. The normalized radial wave functions are
\begin{equation}
  R_{n,m}\left(r\right)=
{\cal N}\frac{1}{\ell}
\left(\frac{r}{\ell}\right)^{\left|m+\Delta\right|}L_n^{|m+\Delta|}\left(\frac{r^2}{2\ell^2}\right)e^{-\frac{r^2}{4\ell^2}}
\label{eq:radial}
\end{equation}
where $\ell\equiv \sqrt{\hbar/\left(qB\right)}$ is the magnetic
length, $L_n^{m}$ are the Laguerre polynomials, and the normalization is ${\cal N}\equiv
\sqrt{n!/\left(\Gamma\left(\left|m+\Delta\right|+n+1\right) 2\pi 2^{|m+\Delta|}\right)}$.
The energy is
\begin{equation}
E_{n,m}=\hbar\omega_{c}\left[n+\frac{1}{2}\left|m+\Delta\right|-\frac{1}{2}\left(m+\Delta\right)+\frac{1}{2}\right]
\end{equation}
where $\omega_c\equiv qB/M$ is the cyclotron frequency.
The spectrum has the required feature that it is invariant under the
transformation $\Delta\to \Delta' = \Delta - 1$, equivalent to the
insertion of a flux quantum at the origin. This amounts to the change
$m \to m' = m+1$. Thus, it is sufficient to study the range $0\leq \Delta <1$ to cover all possible cases.  For positive $ m+\Delta \geq 0 $
the energy depends only on $n$: these sets of states thus form highly
degenerate Landau levels for the relative motion of the anyons.

%\section{Lowest Landau Level Approximation}

We now reintroduce the potential $U\neq 0$. We do this within a
(generalized) lowest Landau level approximation, in which $U$ is taken
only to lift the degeneracy of the lowest energy states.
This corresponds to retaining only those levels
with $n=0$ and $m+\Delta\geq 0$.  Noting that the lowest energy
state lies higher in energy by $\hbar\omega_{c}\left(1-\Delta\right)$,
the lowest Landau level approximation for the relative motion of the anyons is valid for
$U\ell^{2}\ll\hbar\omega_{c}\left(1-\Delta\right)$.
We denote the set of degenerate basis states as $ | j \rangle $ where
$j = 0,1,2,\ldots $, and expand the wave function as 
$|\psi\rangle \equiv \sum \psi_j  |j\rangle$. 
The Schr{\" o}dinger equation becomes
\begin{eqnarray}
\nonumber
\epsilon \psi_{j}   & = & 
\left[\sqrt{\left(j+\Delta-1\right)\left(j+\Delta\right)}\psi_{j-2}\right. \\
 & & + \left. \sqrt{\left(j+\Delta+1\right)\left(j+\Delta+2\right)}\psi_{j+2}\right]
\label{eq:difference}
\end{eqnarray}
for the amplitudes $\psi_j$, where $\epsilon \equiv E/(U\ell^2)$ is
the dimensionless measure of the energy. Note that there are no terms
that couple odd and even values of $j$. Physically, this arises from
the fact that the potential is invariant under spatial
inversion $(x,y)\to (-x,-y)$, and so the parity of the wave function
is a good quantum number.  Thus the Schr\"{o}dinger equation takes the
form of two decoupled sets of difference equations. We write the
general solution in terms of the ``even'' and ``odd'' channels as
\begin{eqnarray} 
|\psi^{\rm e} \rangle & = & \sum_{p=0}^{\infty}\psi^{\rm e}_{2p}|  2p \rangle \\
|\psi^{\rm o} \rangle & = & \sum_{p=0}^{\infty}\psi^{\rm o}_{2p+1}| 2p+1 \rangle
\end{eqnarray}
For distinguishable anyons, the wave function can be a linear
superposition of these two solutions. However, for 
indistinguishable anyons, the boundary condition
(\ref{eq:indistinguishable}) requires that only the even solution 
contributes.

%\section{Scattering Theory}

We shall construct the wave function at large distances from the
origin, $r\gg \ell$. In this limit, contributions are from single
particle states with $j\gg 1$.  For $j \gg \epsilon, 1$, the Schr{\"
o}dinger equation (\ref{eq:difference}) has the wave-like solutions
(normalized to unit density per orbital $j$) \cite{footnotewkb}
\begin{equation}
    \psi_{j} = e^{ i \theta_0 j} \quad 
[ \theta_0 = \pm \pi/4, \pm 3\pi/4]
\label{eq:waves}
\end{equation}
These solutions can be viewed as incoming and outgoing waves in the
discrete semi-infinite one-dimensional system defined by the sites $j
= 0,1,2\ldots$. 
To understand the nature of these solutions, it is useful to construct
their spatial wave functions
\begin{equation}
\left\langle \bm{r}  |  \psi \right\rangle = \sum_{j}  e^{{i\theta_0 j}}e^{-i (j +\Delta)\theta} R_{0,j}\left(r\right) \,.
\end{equation}
At large radius $r$, the wave function has significant amplitude under
the condition that $\theta \simeq \theta_0 = \pm \pi/4, \pm 3\pi/4$.
That is, the wave function is peaked in these four angular
directions. These are the directions along which the zero energy
equipotentials of the electrostatic potential extend. Recalling that,
in the semi-classical approximation, the particle moves along the
equipotentials of the electrostatic potential, one sees that the
derived angles correspond to the two incoming ($-\pi/4, 3\pi/4$) and
the two outgoing ($\pi/4, -3\pi/4$) channels of the saddle-point
potential.

For a general scattering problem on a semi-infinite one-dimensional
system (e.g. waves on a string which is clamped at one end), at a
fixed energy (frequency) one expects there to be only {\it two}
wave-like solutions at large distances; these can be taken to be the
incoming and the outgoing waves.  That, in the present case, there are
{\it four} wave-like solutions is a special feature of the problem,
which arises from the fact that (as above) the Schr{\" o}dinger equation
(\ref{eq:difference}) conserves the parity. That is, the sites with
$j$ even and the sites with $j$ odd each behave as independent
semi-infinite one-dimensional systems.  For each parity (even or odd)
there is one incoming mode and one outgoing mode.  
It
is  convenient to re-express the
four modes (\ref{eq:waves}) in terms of modes of definite parity.
The explicit forms (normalised to unit density per orbital $j$) are
\begin{eqnarray}
\psi^{\rm e/o, in}_j & = & \frac{1}{\sqrt{2}} \left[ e^{-i\frac{\pi}{4} j} \pm e^{i\frac{3\pi}{4} j}\right]\\
\psi^{\rm e/o, out}_j & = &\frac{1}{\sqrt{2}} \left[ e^{i\frac{\pi}{4} j} \pm e^{-i\frac{3\pi}{4} j}\right]
\end{eqnarray}
which are readily verified to have the feature that $\psi^{\rm e}_j$
($\psi^{\rm o}_j$) is non-zero only for $j=$ even (odd). The ``in'' and
``out'' labels are identified by the fact that the states have large
amplitude on the incoming ($\theta = -\pi/4, 3\pi/4$) or outgoing
($\theta = \pi/4, -3\pi/4$) channels of the saddle point potential.

For the semi-infinite one-dimensional scattering problem, the
asymptotic (large distance) incoming and outgoing waves are coupled,
with scattering from incoming to outgoing waves occurring at small
distances.  In the problem of interest here this scattering at small
distances (small $j$) conserves parity, so the incoming mode in the
even (odd) channel can scatter only to the outgoing mode in the even
(odd) channel. By conservation of particle flux, the scattering from
incoming to outgoing modes can amount only to a phase shift.  Hence,
the (unnormalised) energy eigenstates are of the form
\begin{equation}
\psi^{\rm e/o}_{j}  \sim \psi^{\rm e/o, in}_j +
e^{i\zeta^{\rm e/o}\left(\epsilon\right)}\psi^{\rm e/o, out}_j   \quad [j \gg \epsilon, 1]
\label{eq:estate}
\end{equation}
Thus, the energy eigenstates are fully characterized by two scattering phases:
$\zeta^{\rm e}\left(\epsilon\right)$ and $\zeta^{\rm o}\left(\epsilon\right) $
which correspond to the even- and odd-parity wave functions.
   
For indistinguishable anyons (\ref{eq:indistinguishable}) only
$\psi^{\rm e}$ is relevant.  There is only one incoming and one
outgoing channel for the relative co-ordinate, so the scattering is
described only by a single phase shift, $\zeta^{\rm e}(\epsilon)$.

For distinguishable particles, both $\psi^{\rm e}$ and $\psi^{\rm o}$
can contribute.  In this case, it is instructive to disentangle the
above transformation into states of definite parity, and to determine
the {\it transmission probability}. This is defined as the probability for
transmission from a state that is an incoming wave along the definite
angular direction $\theta \simeq -\pi/4$, into a state that is outgoing
along $\theta \simeq +\pi/4$. (These angles match the convention
chosen in Ref.~\cite{FANDH}.)  The incoming wave (normalised to unit
density per orbital) is
\begin{equation}
e^{-i\frac{\pi}{4} j} = \frac{1}{\sqrt{2}}\left[\psi^{\rm e, in}_j + \psi^{\rm o, in}_j\right]
\end{equation}
which is scattered into the state
\begin{equation}
\frac{1}{\sqrt{2}}\left[e^{i\zeta^{\rm e}}\psi^{\rm e, out}_j + e^{i\zeta^{\rm o}}\psi^{\rm o, out}_j\right]
\end{equation}
Noting that the transmitted wave is 
\begin{equation}
e^{+i\frac{\pi}{4} j} = \frac{1}{\sqrt{2}}\left[\psi^{\rm e, out}_j + \psi^{\rm o, out}_j\right]
\end{equation}
and using the fact that $\psi^{\rm e,out}$ and  $\psi^{\rm o,out}$ are orthogonal, one sees that the transmission probability is
\begin{equation}
T=\left|\frac{1}{2}\left(e^{i\zeta^{\rm e}} +e^{i\zeta^{\rm o}}\right)\right|^2 =  \frac{1}{2}\left\{1+\cos\left[\zeta^{\rm e}\left(\epsilon\right)-\zeta^{\rm o}\left(\epsilon\right)\right]\right\}
\label{eq:t}
\end{equation}

%\section{Solution of the Scattering Problem}

We have determined the functions $\zeta^{\rm e/o}(\epsilon)$ --
 thereby solving the scattering problem for the relative co-ordinate
 -- by a numerical construction of the Green's function of the
 discrete Hamiltonian in (\ref{eq:difference}).  We study an
 approximate version of the full model, in which we treat $j=0,N$
 according to the exact Hamiltonian (\ref{eq:difference}), but take
 the hopping matrix elements for $j=N+1,\infty$ to be constant, and
 equal to $\sqrt{\left(N+\Delta-1\right)\left(N+\Delta\right)}$. The
 method becomes increasingly accurate as $N\to \infty$.  Following
 standard techniques\cite{datta}, the region with $j\geq N+1$ can be
 replaced by a self-energy, and the Green's function for $j\leq N$ is:
\begin{equation}
\hat{G}(\epsilon)=\left[\epsilon\hat{I}-\hat{H}-\hat{\sigma}\right]^{-1}
\end{equation}
where the self-energy  has matrix representation
\begin{equation}
\sigma_{ij}=\frac{1}{2}\delta_{i,N}\delta_{j,N}\left(\epsilon-i\sqrt{4\left(N+\Delta+\
1\right)\left(N+\Delta+2\right)-\epsilon^{2}}\right)
\end{equation}
The Green's function for $j\leq  N$ is then found by numerical
inversion of the finite matrix (for large finite $N$).

Using the fact that, for a given element $i$, the Green's function
$G_{i,j}$ for $j<i$ is an energy eigenfunction, we can use this to
compare to the wave function (\ref{eq:estate}) in the asymptotic regime
$j\to \infty$.  Using (\ref{eq:estate}), the scattering phase
can be extracted from the ratios
\begin{eqnarray}
\frac{\psi^{\rm e}_{4p+2}}{\psi^{\rm e}_{4p}} & \stackrel{p\rightarrow\infty}{\longrightarrow}  & \tan\left[\frac{-\zeta^{\rm e}\left(\epsilon\right) + \frac{\epsilon}{2}\ln(4p)}{2}\right] \\
\frac{\psi^{\rm o}_{4p+3}}{\psi^{\rm o}_{4p+1}} & \stackrel{p\rightarrow\infty}{\longrightarrow}  & \tan\left[\frac{-\zeta^{\rm o}\left(\epsilon\right)-\pi/2 + \frac{\epsilon}{2}\ln(4p)}{2}\right] 
\end{eqnarray}
where the logarithms on the right-hand side follow from the
corrections to the wave functions (\ref{eq:waves}) described in
\cite{footnotewkb}.  
In this way we
can numerically construct the scattering phases
$\zeta^{\rm e/o}\left(\epsilon\right) $ and from these the transmission
probability $T$ (\ref{eq:t}).
The numerical results converge rapidly with increasing $N$, becoming
independent of system size for $N\gtrsim 100$. We show results for
$N=2000$.

\begin{figure}      
\centering
\includegraphics[angle=0,scale=0.35]{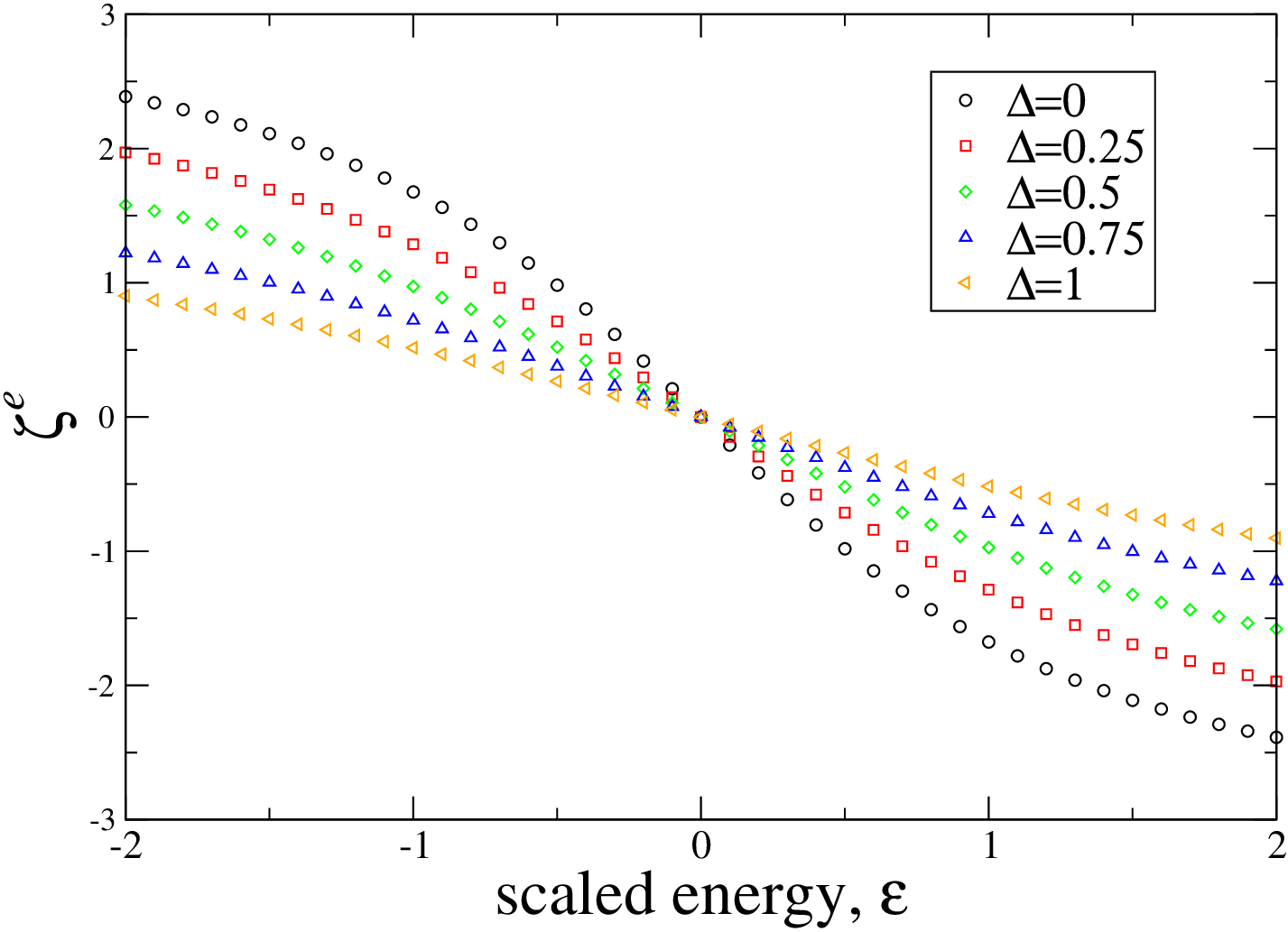}
\caption{(Colour online.) Results of numerical calculations of the
scattering phase shift for the relative motion of two anyons in the
even parity channel, $\zeta^{\rm e}(\epsilon)$, as a function of
dimensionless energy $\epsilon = E/U\ell^2$ at several values of the
anyonic statistics parameter $\Delta$.  }
\label{EVEN}
\end{figure}
\begin{figure}      
\centering \includegraphics[angle=0,scale=0.35]{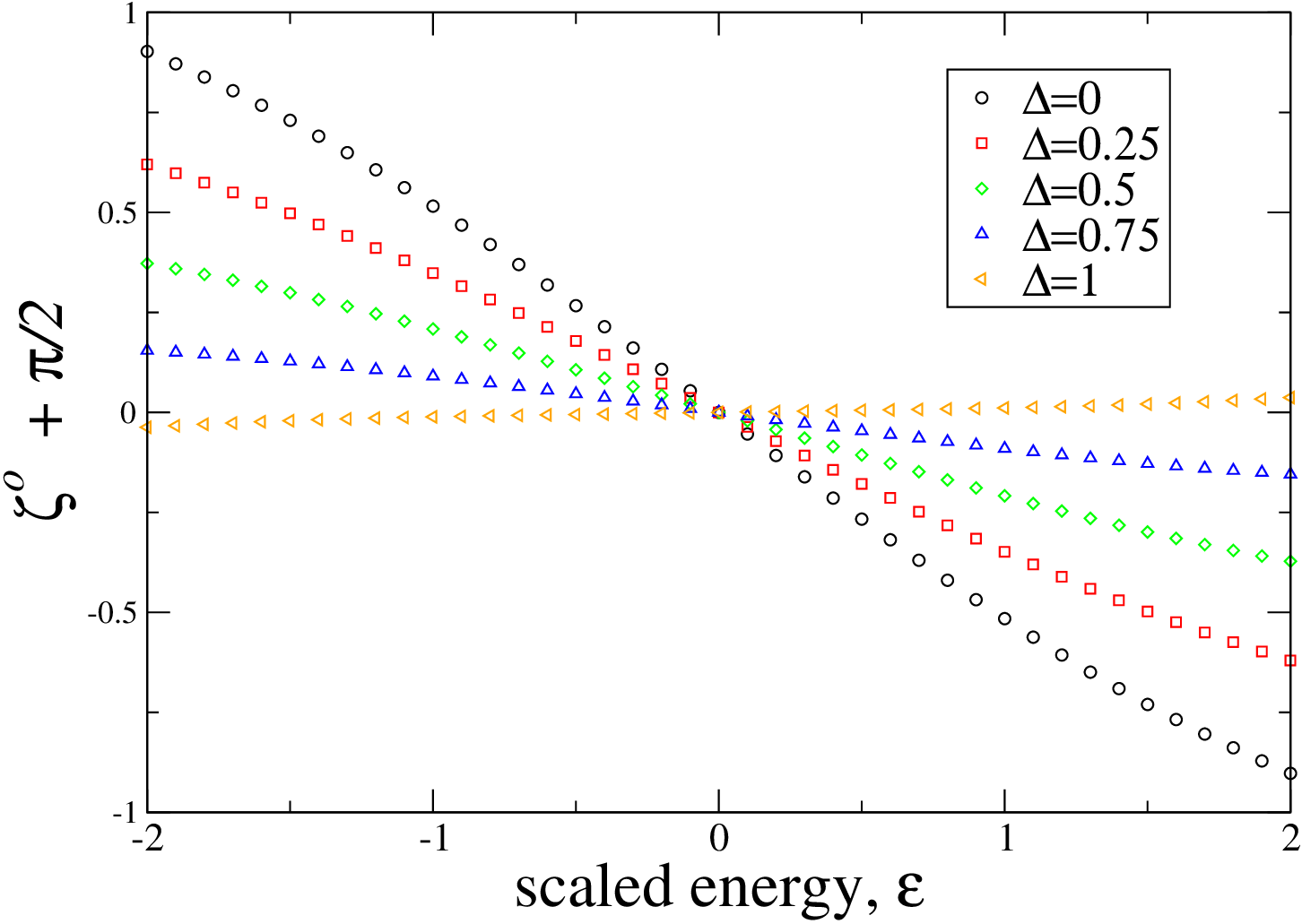}
\caption{(Colour online.) Same as Fig.~\ref{EVEN}, but for 
the
 odd parity channel, $\zeta^{\rm o}(\epsilon)$.}
\label{ODD}
\end{figure}
Figs.~\ref{EVEN} and~\ref{ODD} show our numerical results for the
scattering phase shifts for the even and odd parity channels
respectively, as a function of energy for several values of $\Delta$.
These two functions, over the range $0\leq \Delta <1$, fully describe
the scattering properties of the relative co-ordinate of quantum Hall
anyons in the lowest Landau level.  The results shown for $\Delta =1$
correspond to the case $\Delta = 1-$ in which the state $m=-1$ remains
excluded from contribution to the lowest Landau level.  In this case,
the spectrum is identical to that for $\Delta' =0$ and $m' = m+1$, but
with the removal of the state at $m'=0$. 
From (\ref{eq:estate}), and taking account of a $\pi/2$ phase shift arising from the change $m'=m+1$, 
one expects
$\zeta^{\rm e}_{\Delta = 1-}(\epsilon) = \zeta^{\rm o}_{\Delta' = 0}(\epsilon)+\pi/2$,
which is indeed found to hold to high accuracy in the numerical
results. 
\begin{figure}      
\centering
\includegraphics[angle=0,scale=0.35]{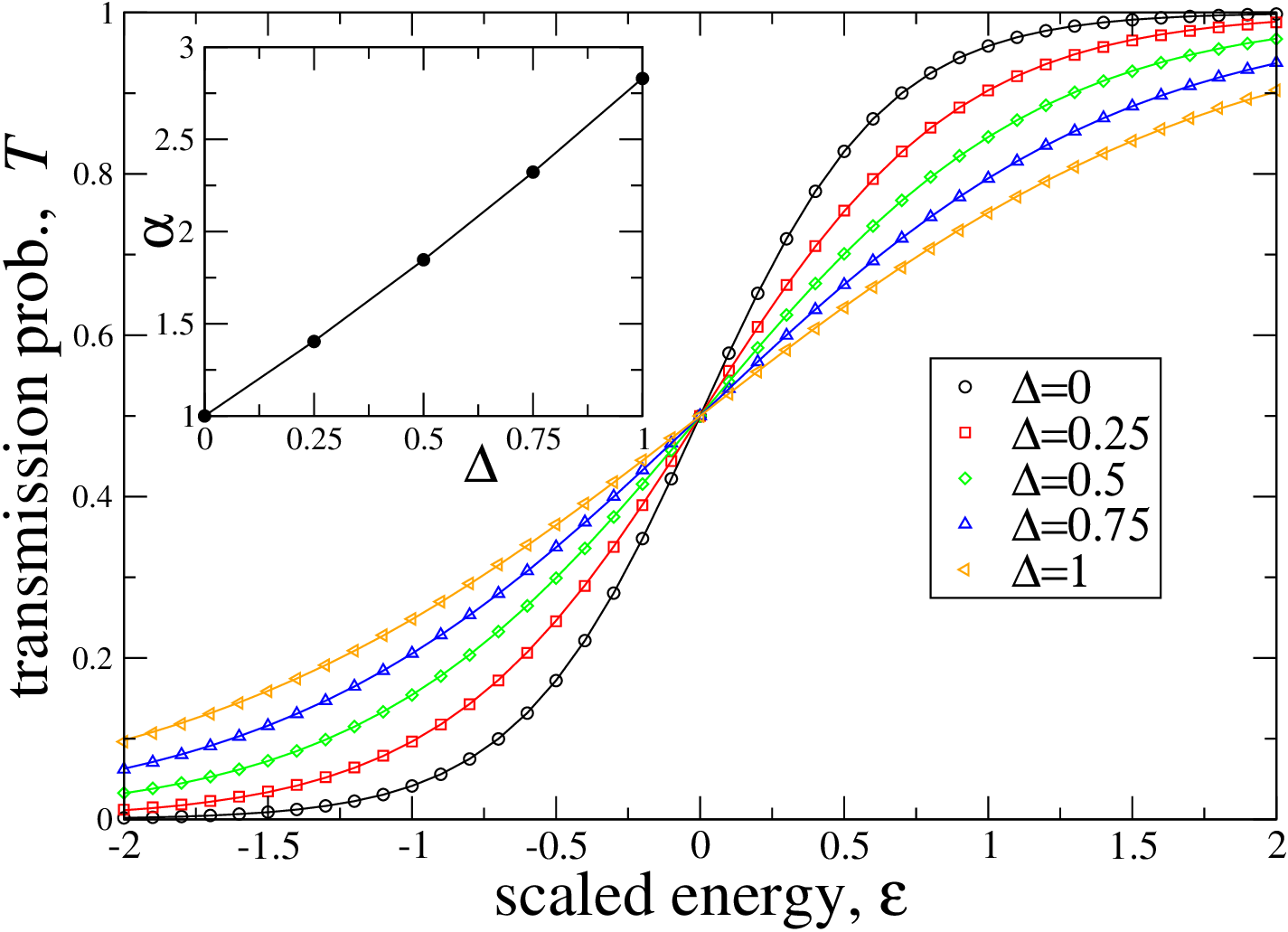}
\caption{(Colour online.) Transmission probability $ T $ for the
relative coordinate of two quantum Hall anyons, as a function of the
scaled energy $\epsilon\equiv E_r/(U\ell^2)$.  The points are results
of numerical calculations for several values of the anyon statistics
parameter $\Delta$. The lines are fits, using the function
(\ref{eq:fit}) with parameter $\alpha$ chosen as shown in the
inset. For $\Delta =0$ choosing $\alpha=1$ gives the exact analytic
result\protect\cite{FANDH}; in this case, the comparison with
numerical results illustrates the accuracy of our numerical approach
at this system size ($N=2000$).}
\label{fig:transmission}
\end{figure}
In Fig.~\ref{fig:transmission} we have used these results to 
determine the
transmission coefficient (\ref{eq:t}).  For $\Delta =0$, the results accurately reproduce the exact
analytical solution\cite{FANDH}, showing that our method is working
correctly and is well converged.  For $\Delta \neq 0$, the results of
Ref.\cite{FANDH} do not apply, and no analytic solution is available.  As
compared to the case $\Delta =0$, the effect of increasing $\Delta$ is
a broadening of the width in energy over which the transmission
coefficient rises from $0$ ($\epsilon\ll -1$) to $1$ ($\epsilon \gg
1$).  Thus, our results show that increasing $ \Delta $ leads to an
{\it increase} of the tunnelling rate through the saddle point. Indeed,
we find that, to a very good approximation, the results can be fitted by the
function
\begin{equation}
\label{eq:fit}
T(\epsilon) = \frac{1}{1+\exp\left({-\frac{\pi\epsilon}{\alpha}}\right)}
\end{equation}
with the fit parameter $\alpha$ given approximately by $\alpha = 1 +
1.55378\Delta + 0.277179\Delta^2$. (See the inset to
Fig.~\ref{fig:transmission}.) 

We note that our results apply for the case of two anyons in a
symmetrical saddle point potential (\ref{eq:ham}). The more general
case can be considered by noting that $U_y y^2 - U_x x^2 =
\frac{1}{2}(U_y+U_x) \left(y^2 - x^2\right) + \frac{1}{2}(U_y-U_x)
\bm{r}^2$, leading to an additional central (rotationally invariant)
term $\propto \bm{r}^2$. This term modifies the Schr{\" o}dinger
equation (\ref{eq:difference}) and could lead to a change in the phase
shifts and transmission probabilities. In the same way, central
(rotationally invariant) anyon-anyon interactions could be included
within the same formalism. The solution of these more general cases is
beyond the scope of the present paper, so the influence of these
perturbations on the scattering properties 
remains an open question.

In summary, we have provided a solution of the scattering problem for
two anyons in a quadratic saddle point potential and
perpendicular magnetic field, through separation into centre-of-mass
and relative coordinates. The scattering theory for the centre-of-mass
coordinate has previously been solved analytically\cite{FANDH}. The
scattering for the relative co-ordinate is characterized by two
energy-dependent phase shifts (for even and odd parities). We have
computed these phase shifts within a lowest Landau level
approximation. Our results provide a complete solution of the two-anyon scattering problem. They show that the two-particle scattering
properties in the vicinity of a point contact depend on the anyonic
statistics parameter. This shows that one can hope to obtain
experimental signatures of anyonic statistics in point-contact
devices.  We believe that the approach and solution we have described
provide a useful basis on which to build further theoretical studies
of non-equilibrium properties of anyons -- under conditions of bias
and/or ``beam dilution''\cite{comforti} where multiple quasiparticles
may enter the point contact region. Depending on the experimental
set-up, and the relevant observables, the two-particle scattering
problem in the co-ordinates $\bm{r}_1, \bm{r}_2$ should be decomposed into
relative and centre-of-mass co-ordinates for which the scattering
properties follow the results presented.

\acknowledgments{NRC acknowledges the support of EPSRC GR/S61263/01.}

%\bibliography{references}

\begin{thebibliography}{10}
\bibitem{footnote} We note that two distinguishable particles which
exhibit mutual anyon statistics -- e.g. quasiparticles of two
different quantum Hall fluids -- cannot be viewed as non-interacting,
except for the cases $\Delta =0,1$ for which the boundary condition
(\ref{eq:anyon}) becomes trivial.
\end{thebibliography}
%\bibliographystyle{prsty}

\end{document}